\documentclass[aps,prr,twocolumn,superscriptaddress,showpacs,floatfix]{revtex4-1}
\usepackage{color}
\usepackage{dcolumn}   % needed for some tables
\usepackage{bm}        % for math
\usepackage{amssymb}   % for math
\usepackage{amsmath}
\usepackage{gensymb}   % for \degree symbol
\usepackage{bbm}
\usepackage{hyperref}
\usepackage{graphicx}
\usepackage{array}
\usepackage{braket}
\usepackage{multirow}
\usepackage{natbib}
\usepackage{overpic}

\newcommand{\be}{\begin{equation}}
\newcommand{\ee}{\end{equation}}

\newcommand{\br}{{{\bf{r}}}}

\newcommand{\bea}{\begin{eqnarray}}
\newcommand{\eea}{\end{eqnarray}}
\newcommand{\beal}{\begin{align}}
\newcommand{\eeal}{\end{align}}
\newcommand{\ra}{\rangle}
\newcommand{\la}{\langle}

\newcommand{\pdg}{{\phantom\dagger}}

\newcommand{\fQ}{F${\cal Q}$}

\begin{document}

\title{Thermal and field-induced transitions in ferroquadrupolar Kondo systems}
%\title{Ferroquadrupolar order and field-induced transitions in Kondo systems: Possible application to PrTi$_2$Al$_{20}$}
\author{Frederic Freyer}
\affiliation{Institute for Theoretical Physics, University of Cologne, 50937 Cologne, Germany}
\author{SungBin Lee}
\affiliation{Department of Physics, Korea Advanced Institute of Science and Technology, Daejeon, 34141, Korea}
\author{Yong Baek Kim}
\affiliation{Department of Physics, University of Toronto, Toronto,
Ontario M5S 1A7, Canada}
\author{Simon Trebst}
\affiliation{Institute for Theoretical Physics, University of Cologne, 50937 Cologne, Germany}
\author{Arun Paramekanti}
\email{arunp@physics.utoronto.ca}
\affiliation{Department of Physics, University of Toronto, Toronto,
  Ontario M5S 1A7, Canada}

\begin{abstract}
Recent experiments have examined the impact of a magnetic field on ferroquadrupolar orders in the intermetallic Kondo material PrTi$_2$Al$_{20}$.
Motivated by this, we use extensive Monte Carlo simulations to
study a diamond lattice XY model of non-Kramers pseudospin-$1/2$ Pr$^{3+}$ moments which crucially incorporates three-spin interactions. This model
supports a thermal $Z_3$ Potts ordering transition upon cooling from the paramagnetic phase into the ferroquadrupolar phase.
An applied magnetic field along the [110] direction leads to a thermal Ising transition out of the quadrupolar ordered phase. A magnetic field along the
[001] direction leads to only thermal crossovers, but supports a spinodal transition out of metastable domains which could be strongly pinned
by coupling to elastic lattice deformations.  We propose noise measurements as a potential probe to ``hear'' the spinodal transition.
Our work highlights the importance of multispin interactions in Kondo materials near the small-to-large Fermi surface
transition.
\end{abstract}
\maketitle

\section{ Introduction}

The famous Doniach picture of Kondo lattice compounds suggests a scenario for the small-to-large Fermi surface (FS) transition in Kondo lattice materials. In this framework,
weak Kondo coupling leads to two-spin RKKY interactions which drive rare-earth
local moment ordering and a small FS, while strong Kondo coupling leads to the local moments hybridizing with the conduction electrons resulting in a
heavy Fermi liquid with a large FS \cite{doniach1977kondo,ruderman1954indirect, kasuya1956theory, yosida1957magnetic,stewart1984heavy, lohneysen2007fermi}.
While there has been important work in understanding this physics for materials with local dipole moments \cite{yosida1957magnetic,stewart1984heavy,fisk1995physics,coleman2001fermi,gegenwart2008quantum,si2010heavy},
there is considerably less understanding of higher multipolar orders \cite{morin1982magnetic,cox1987quadrupolar,cox1999exotic,kitagawa1996possible,caciuffo2003multipolar,suzuki2005quadrupolar,kuramoto2009multipole,lee2015optical},

Recently, there has been significant experimental progress in unveiling the rich phase diagram of the cubic rare-earth intermetallics
Pr(TM)$_2$Al$_{20}$ (TM=Ti,V) and PrIr$_2$Zn$_{20}$ \cite{sakai2011kondo,koseki2011ultrasonic,sakai2012thermal,sato2012ferroquadrupolar,onimaru2016exotic,onimaru2011antiferroquadrupolar,sakai2011kondo,shimura2013evidence,onimaru2012simultaneous,onimaru2010superconductivity,sakai2012superconductivity,matsubayashi2012pressure,matsubayashi2014heavy,tsujimoto2014heavy,iwasa2017evidence,taniguchi2016nmr,Kusunose_JPhys2015,Onimaru_PRB2016,Gegenwart_PRB2019} which feature Pr$^{3+}$ local moments coupled to conduction electrons
\cite{ruderman1954indirect, kasuya1956theory, yosida1957magnetic,stewart1984heavy, lohneysen2007fermi}.
The complex multipolar orderings and superconductivity in these compounds may be tuned by the choice of transition metal ion or pressure.
Understanding the broken symmetry states and phase transitions in such multipolar Kondo materials remains a largely open issue.

One basic question which arises when one confronts the plethora of broken symmetry states in Kondo materials is whether one needs to go beyond
the simple two-spin RKKY model in modelling the effective interaction between local moments. Indeed, as the Kondo coupling in heavy fermion materials
increases, we expect multispin interactions can arise from higher-order perturbation theory beyond the simple RKKY limit. One setting in which
such multispin interactions have been investigated extensively is in the vicinity of Mott transitions in quasi-two-dimensional organic
materials \cite{Motrunich2005,Motrunich2007,sheng2008boson,sheng2009spin,grover2010weak} where it has been shown to
potentially stabilize exotic quantum spin liquids.
From this viewpoint, we expect multispin interactions to also emerge naturally in Kondo materials if we view the
the large-to-small FS transition as an ``orbital selective Mott transition'' of the $f$-electrons
\cite{de2005orbital,de2009orbital}. The impact of such couplings has only recently been investigated
in multipolar Kondo systems \cite{freyer2018two,lee2018landau,Patri2019,Patri2020}, although there has been some suggestive previous work in
dipolar Kondo materials \cite{Staunton2017,akagi2012hidden}. Given this, we ask the following questions.
Are there any
heavy fermion multipolar systems where multispin couplings play a role? Can such interactions lead to
observable signatures?

We address these questions in the context of recent experiments on the Pr(TM)$_2$X$_{20}$ family of materials, where the Pr$^{3+}$ ions feature
a non-Kramers ground state doublet, which acts as a pseudospin-$1/2$ degree of freedom on the diamond lattice \cite{sato2012ferroquadrupolar,onimaru2016exotic}.
As discussed in the literature, two components of this pseudospin
carry a quadrupolar moment while the third component describes an
octupolar moment \cite{onimaru2016exotic,shiina1997magnetic}. In this paper, we focus on PrTi$_2$Al$_{20}$, which has been proposed to host a
ferroquadrupolar (\fQ) ordered ground state \cite{sakai2011kondo,sakai2012superconductivity,koseki2011ultrasonic,sato2012ferroquadrupolar,taniguchi2016nmr}
below $T_Q \approx 2$K,
well before the system enters a low temperature superconducting state with transition temperature $T_{\rm SC} \approx 0.2$K.

Recent
experiments have studied the non-trivial impact of a magnetic field on this ferroquadrupolar ordered state, discovering a strong
dependence of the response on the field direction \cite{taniguchi2016nmr,taniguchi2019field,kittaka2019field}.
For a magnetic field along the [111] direction, the quadrupolar transition appears nearly
unaffected, while there appear to be distinct field-induced transitions for fields along [001] and [110] directions.
We argue here that an appropriate low-energy microscopic
model for this material must necessarily include three-spin interactions, and that it reveals itself via the impact of a magnetic field.

Our key results are the following. We show that the model pseudospin-$1/2$ Hamiltonian for local Pr$^{3+}$ moments
must include crucial symmetry-allowed three-spin couplings. We use classical Monte Carlo (MC) simulations to study the ordered
states, thermal fluctuations, and the impact of a magnetic field in this model. We uncover
thermal and field-induced phase transitions and crossovers which are qualitatively consistent with experimental observations. However, our model
does not display a subset of field-induced transitions which have been inferred from certain experiments \cite{taniguchi2019field}. We argue here that such experiments could potentially
probe spinodal transitions out of metastable ground states; such metastable states do exist in the model and may be rendered visible by the strong coupling
between the quadrupolar order and elastic lattice deformations. We propose that noise measurements could be used to ``hear'' such
spinodal transitions. While our work here focusses on PrTi$_2$Al$_{20}$, our main results are broadly applicable to ferro-quadrupolar orders in
diverse materials.

\section{Model}
\begin{figure}[t]
	\includegraphics[width=.75\columnwidth]{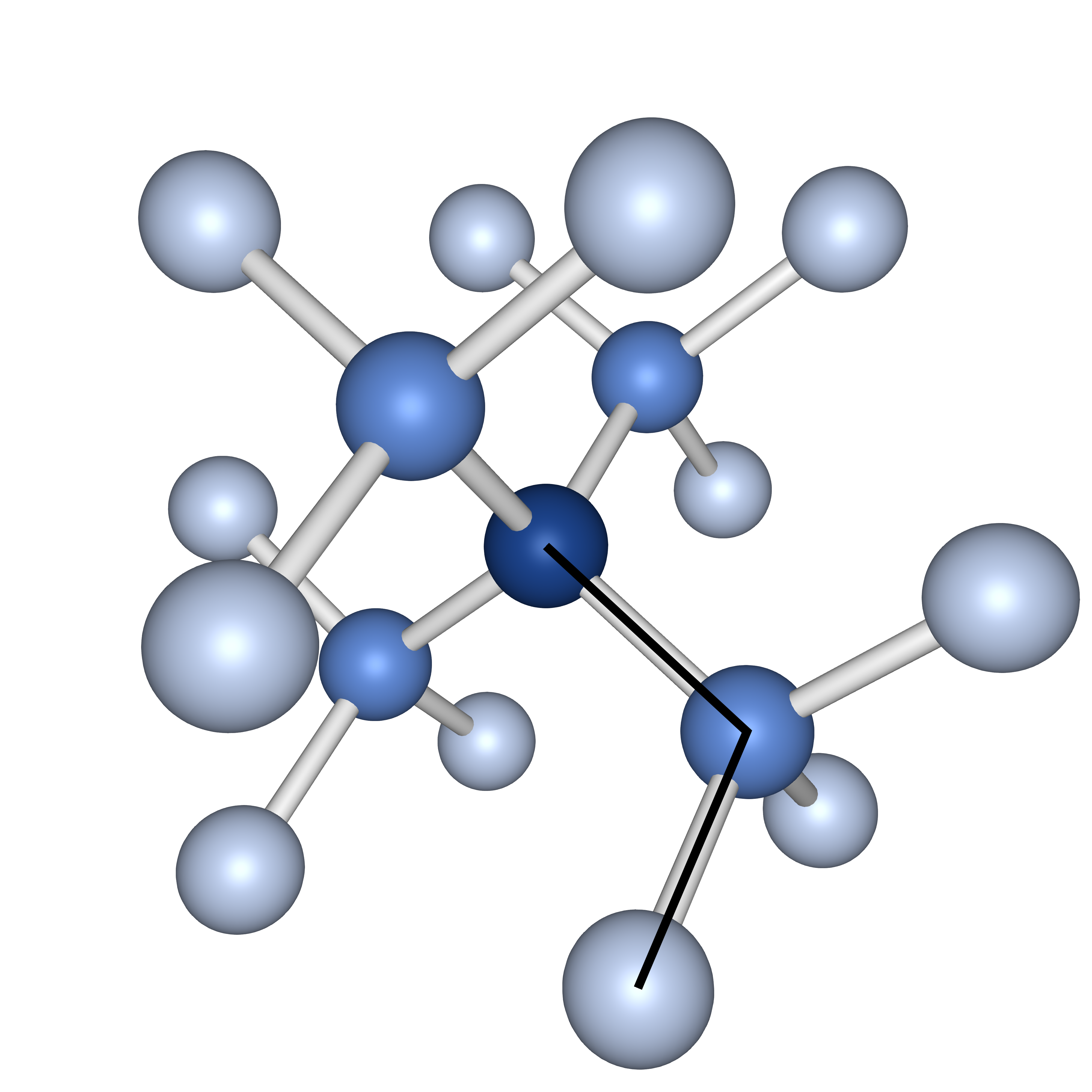}
	\caption{
		Cutout of a diamond lattice with nearest neighbor bonds ($J_1$) drawn in white. The three site triplets of the $\Gamma$ term are constructed by two adjacent nearest neighbor bonds as exemplified by the black line.
		%Diamond lattice with nearest (red, J$_1$) and next nearest neighbors. Three site triplets originating from the central site are given by a red and a yellow bond. An example of this is given by the white dotted line.
	}
	\label{fig:diamond}
\end{figure}

We consider a simple low energy diamond lattice model for the pseudospin-$1/2$ non-Kramers doublets, ignoring higher
crystal field levels which are split off by a relatively big energy scale $\sim 50$\,K, which is much larger
than the observed ferroquadrupolar transition temperature. The pseudospin Hamiltonian we propose takes the form
\bea
\!\!\! H_0 \!&=&\! - J_1 \sum_{\la i j \ra} \! \vec \tau^\perp_i \cdot \vec \tau^\perp_j  \!-\!
i \frac{\Gamma}{2} \! \sum_{\la i j k \ra} (\tau^+_i \tau^+_j \tau^+_k \!-\! {\rm H.c.}) \nonumber \\
\!&\!-\!&\! \alpha \! \sum_i \left[\! \sqrt{3} (B^2_x-B^2_y) \tau^x_i \!+\! (3 B_z^2-B^2) \tau^y_i \!\right] \,,
\label{eq:H0}
\eea
where $\vec\tau^\perp \equiv (\tau_x,\tau_y)$ denotes pseudospin-$1/2$ Pauli matrices and ${\rm H.c.}$ refers to the Hermitian conjugate.
We denote nearest neighbor pairs by $\la ij\ra$, while the notation $\la ijk\ra$ refers to
shortest site-triplets on the diamond lattice as illustrated in Fig.~\ref{fig:diamond}. Our notation for the spin operators follows
Refs.~\onlinecite{freyer2018two,lee2018landau}, and differs from that used in some of the literature \cite{hattori2014antiferro,taniguchi2019field,kittaka2019field}.
In our convention, $\la \tau^x \ra$ corresponds to
$O_{22}$ order, while $\la \tau^y \ra$ refers to $O_{20}$ order, where $O_{22} \propto \sqrt{3}({\cal J}_x^2-{\cal J}_y^2) $ and $O_{20} \propto (3 {\cal J}_z^2 - {\cal J}^2)$
are the standard Steven's operators written in terms of the total angular momentum $\vec {\cal J}$ of the Pr$^{3+}$ ion.

An easy-plane interaction with $J \!\!>\!\! 0$ is appropriate to describe ferroquadrupolar $XY$ order in PrTi$_2$Al$_{20}$.
The magnetic field $\boldsymbol{B}=(B_x,B_y,B_z)$ couples to the pseudospin at ${\cal O}(B^2)$. This arises within second-order perturbation theory \cite{hattori2014antiferro,hattori2016antiferro}  
via intermediate states involving
higher crystal field multiplets, with $\alpha \!>\! 0$.
Most importantly, the term $\Gamma \!>\! 0$ is the simplest symmetry allowed three-spin interaction which breaks the XY symmetry and  leads to a $Z_3$ clock anisotropy.
While such clock terms have been previously discussed within Landau theory 
\cite{hattori2014antiferro,hattori2016antiferro,lee2018landau}, there can be {\it no} such single-site clock anisotropy
term for pseudospin-$1/2$ models. Hence, the clock-like anisotropy for pseudospin-$1/2$ cases
must necessarily arise from multi-site couplings at the lattice scale.
We note that this multispin interaction allows for the
$\tau_z$ eigenvalue to change in steps of $\pm 3$, which {\it cannot arise in any RKKY-type two-spin exchange model}.

Our motivation here is to understand the ordered phases
and thermal transitions of such quadrupolar spin models. We will thus focus on a mean-field theory and large scale classical MC simulations of this model
replacing $\vec\tau^\perp$ by a classical XY vector spin. It would be interesting in the future to examine the impact of quantum spin fluctuations in this model.

\section{Mean-field theory}

\begin{figure}
	\includegraphics[width=\columnwidth]{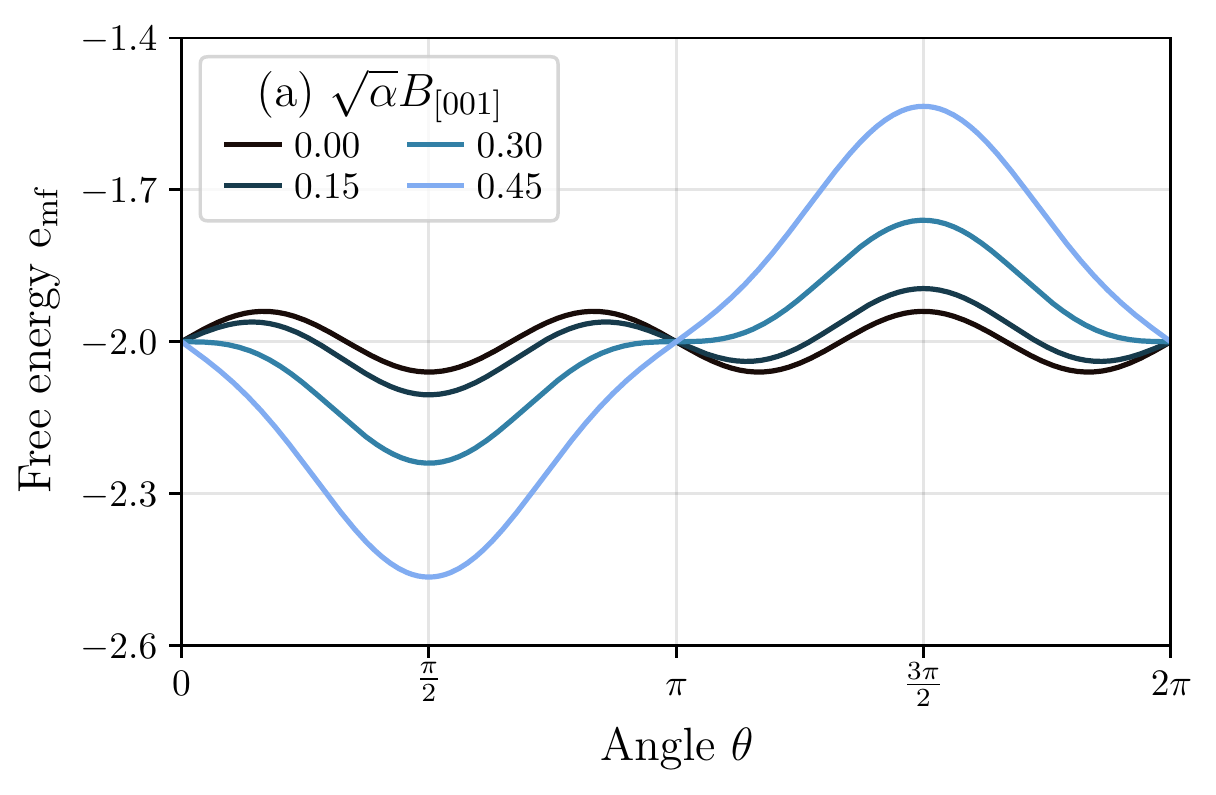}
	\includegraphics[width=\columnwidth]{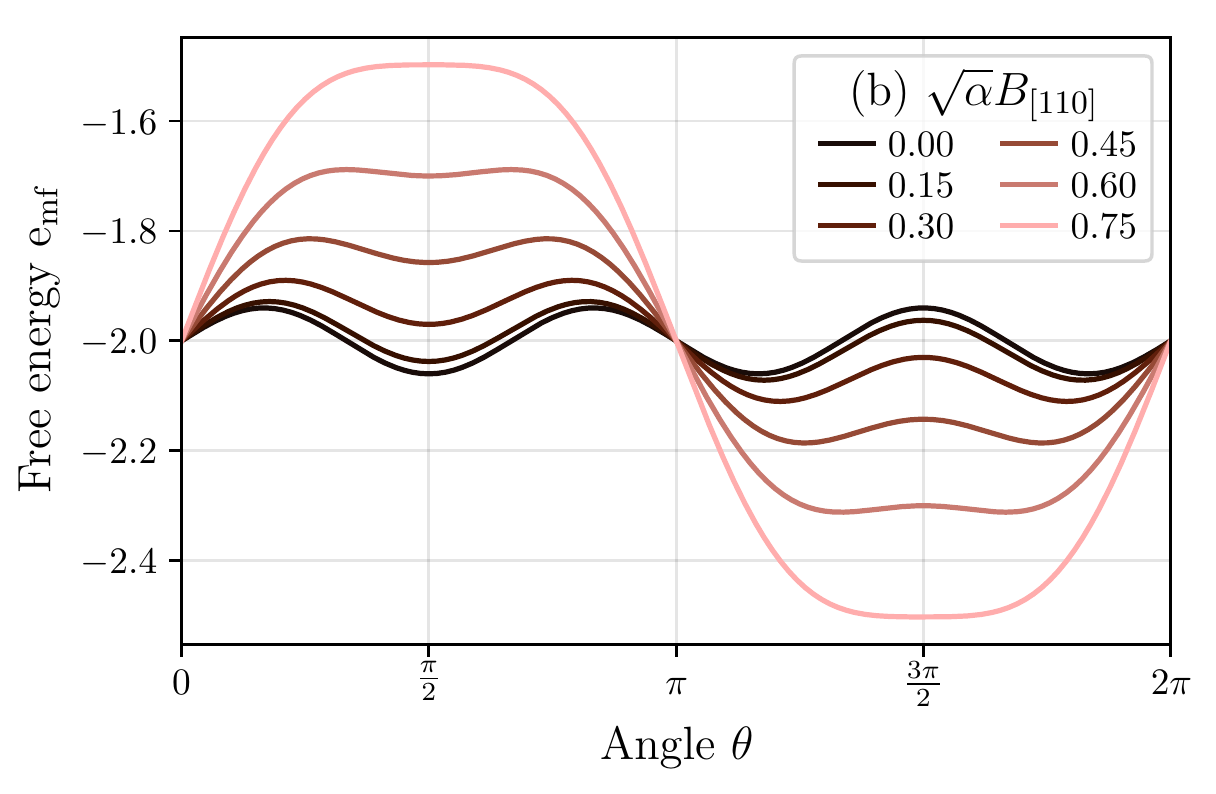}
	\caption{Sketch of the mean field free energy per spin, Eq.~\eqref{eq:MFenergy}, for $J_1\!=\!1$ and $\Gamma\!=\!0.01 J_1$,
	plotted versus $\theta$ for increasing magnetic field along the (a) [001] and (b) [110] directions.	\label{fig:MFenergy}}
\end{figure}

At zero temperature and at mean-field level, we replace uniformly $\tau^+_i = {\rm e}^{i \theta}$ which leads to an energy per spin
\bea
e_{\rm mf} &=& -2 J_1 + 6 \Gamma \sin3\theta \notag \\
&-& \alpha \left[\sqrt{3} (B^2_x - B^2_y) \cos\theta + (3 B_z^2 - B^2) \sin\theta \right] \,.
\label{eq:MFenergy}
\eea
The magnetic field thus competes with the $\Gamma$-term, allowing us to probe the impact
of the reduction of symmetry from $U(1)$ to $Z_3$.
We see that applying a field in the [111] direction will not couple at all to the quadrupolar field. A magnetic field along
$[110]$ direction gives $ \alpha B^2 \sin\theta$, while a field along the $[ 001]$ direction gives $-2 \alpha B^2 \sin\theta$.
Fig.~\ref{fig:MFenergy} plots the energy  landscape as a function of $\theta$ and $B$ for these latter two field
directions.

As seen from Fig.~\ref{fig:MFenergy}(a), a magnetic field along $[001]$ favors $\theta=\pi/2$, while the two other zero-field minima become metastable minima at
$\theta = (7\pi/6-\delta, 11\pi/6+\delta)$, where
$\delta \approx \sqrt{3} \alpha B^2/(54\Gamma-\alpha B^2)$
for weak fields. These metastable minima lie at an energy $\approx \! 3 \alpha B^2$ above the ground state,
vanishing at a field $B_{\rm sp} = \sqrt{9 \Gamma/\alpha}$ which marks a field-induced spinodal transition.

For a $[110]$ field, as seen from Fig.~\ref{fig:MFenergy}(b), there are two energy minima which lie at $\theta = (7\pi/6+\delta', 11\pi/6-\delta')$, where we find that
$\delta' \approx \sqrt{3}\alpha B^2/(108 \Gamma+ \alpha B^2) $
for small $B$. The third zero-field minimum becomes a metastable minimum at $\theta=\pi/2$, which lies at an energy $\approx 3 \alpha B^2/2$ above
the global minima. This will convert the thermal $Z_3$ clock transition into an Ising transition
since the three ground states of the $Z_3$ clock model have been reduced to just two degenerate ground states. Eventually, the two
minima merge at $B^{\star}$ which marks the end point of the Ising transition,
where $B^{\star} = \sqrt{54\Gamma/\alpha}$. The metastable minimum at $\theta=\pi/2$ persists until $B^{\star}$.

We thus expect that for the [001] field direction, the field should immediately round off the $Z_3$ thermal transition into a crossover by
selecting one of the three ground states. For the [110] direction on the other hand, we expect the $Z_3$ thermal transition to convert
into an Ising transition for arbitrarily weak fields, with the Ising transition vanishing above a critical field $B^{\star}$.

\section{Metastable minima}

The reason why the metastable minima in Fig.~\ref{fig:MFenergy} may be important to explore in this system is the following. Imagine we consider starting from a
paramagnetic state at high temperature. When we cool below the ferroquadrupolar transition
$T_Q$ at zero field, we would end up
having different $Z_3$ domains of a typical size $L(T)$ at some temperature $T < T_c$. 
Ferroquadrupolar order in this system induces a lattice distortion, which arises from coupling an elastic distortion field $\vec\varphi$ to the
quadrupolar degree of freedom, which can be understood via a coarse-grained Hamiltonian
\bea
\!\!\! H \!&=&\! H_0 + \frac{1}{2} \kappa \int d^3\br \vec\varphi(\br)\cdot\vec\varphi(\br) - \lambda \int d^3\br \vec\varphi (\br) \cdot \vec\tau^\perp(\br)\,,\,\,\,\,
\eea
where $\vec\tau^\perp(\br)$ is the coarse grained quadrupolar order parameter, and $\lambda$ denotes the magnetoelastic coupling.
The two-component elastic distortion field $\vec\varphi$ may be written in terms of the elastic strain tensor $\varepsilon$, as
$\varphi_x =\varepsilon_{xx}-\varepsilon_{yy}$ and $\varphi_y = (2 \varepsilon_{zz}-\varepsilon_{xx}-\varepsilon_{yy})/\sqrt{3}$.
The impact of quadrupolar order will thus be to produce a small nonzero lattice distortion $\vec\varphi$.
This elastic deformation along different directions in the different domains
will tend to collectively pin the local order. Thus, we see that while an applied $[001]$ field will favor a single domain, we have to thermally
excite the system out of the metastable domains in order to get to the true equilibrium state. If thermal fluctuations are not significant
at low temperature, then such domains might get stuck until we reach a threshold field corresponding to a mean-field spinodal transition; this
effect may reveal itself in certain experiments.

%\begin{figure}
%	%\includegraphics[width=\columnwidth]{pos_hy_free_energy_sketch}
%	\caption{Sketch of the free energy for increasing magnetic field in [001] direction.}
%\end{figure}

\section{Monte Carlo simulations}

% order parameter in context of update:
% \vec{M} = < < \vec{S} >_sites >_MC
% is the naive magnetization - (the MC average of) the average spin direction per site (2 component vector)

% order parameter m_FQ, only used in context with fig 3b)
% M_FQ = < | < \vec{S] }>_sites | >_MC
% is a number corresponding to the average amplitude of the magnetization, ignoring direction

% If all spins collectively point in the same direction (S_i = S_0 for all i), but that direction is free to rotate (S_0 is any normalized vector), then
% \vec{M} \approx \vec{0} (assuming MC doesn't get stuck)
% M_FQ \approx 1

We have carried out extensive classical MC simulations of the Hamiltonian $H_0$ from Eq.~\eqref{eq:H0}. 
While standard MC updates sufficed to explore the equilibrium phase diagram via measurements of the specific heat and ferroquadrupolar XY order parameter 
$\vec M_{\rm FQ} \!=\! \sum_i \vec \tau_i^\perp$,
exploring the metastable transitions required us to choose a special update engineered to probe the free energy as function of  the
angle $\theta$ of $\vec M_{\rm FQ}$.
The update involves a local update conserving the direction of $\vec M_{\rm FQ}$,
and a global update jumping between two angles.
Combining multiple such simulations at slowly varying angles (typically $\Delta\theta \!=\! 2\pi/1080$) we recover the
relative weights between them, and ultimately estimate the free energy landscape. Further details on this procedure are provided in the Appendix.
Simulations were typically done with two million thermalization and eight million measurement sweeps
for a linear system size of $L \!=\! 9$ (corresponding to $2L^3 \!=\! 1458$ spins) in mapping out the phase diagram and  
for $L \!=\! 6$ when studying the metastable regions.

\subsection{Zero field phase diagram}
\begin{figure}
	\includegraphics[width=\columnwidth]{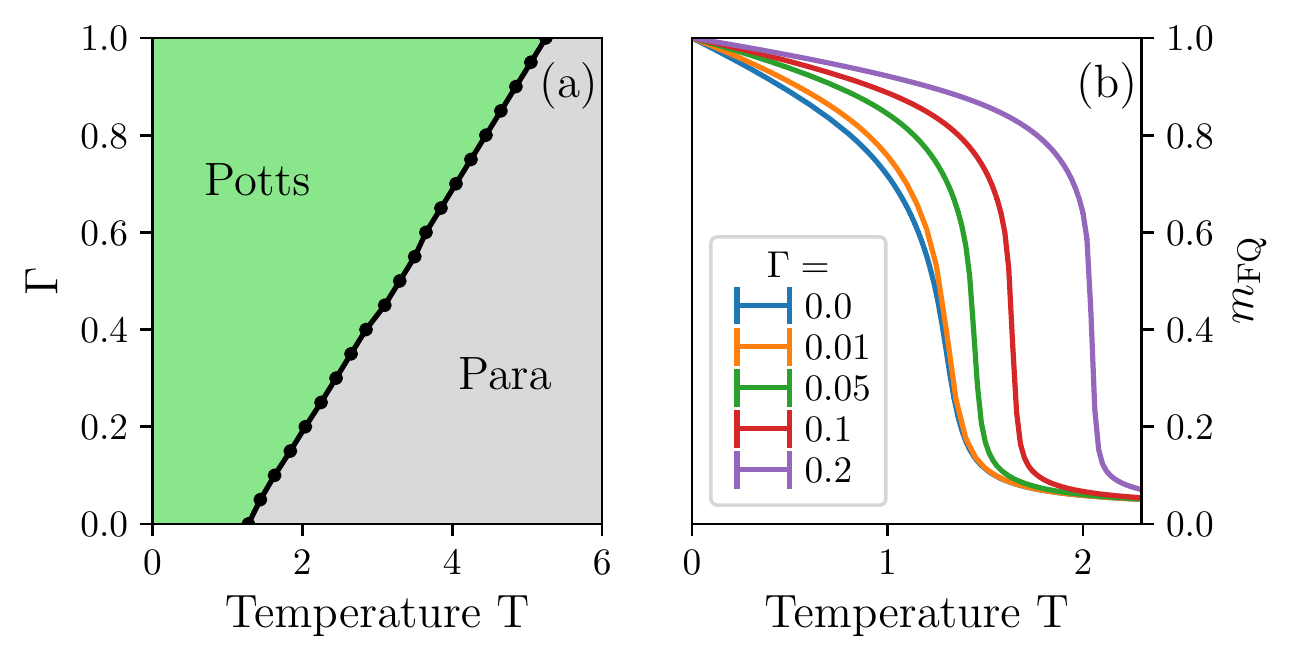}
	\caption{(a) Zero field phase diagram of Hamiltonian $H_0$, with $J_1 = 1$, as a function of temperature $T$ and varying clock anisotropy $\Gamma$.
	(b) Ferroquadrupolar order parameter $m_{\rm FQ} \!=\! \la |\vec M_{\rm FQ}| \ra/N$ 
	as a function of $T$ for horizontal cuts through the phase diagram in panel (a). $m_{\rm FQ}$
	vanishes continuously for $\Gamma=0$, but becomes discontinuous for $\Gamma\neq 0$; this discontinuity becomes more pronounced for large $\Gamma$.}
	\label{gamma_T_fig}
\end{figure}

The phase diagram of the model with $J_1=1$ in the absence of any magnetic field is shown in Fig.~\ref{gamma_T_fig}(a).
Based on a finite-size scaling analysis of specific heat data (for $L = 6, \ldots, 12$) we find a sharp thermodynamic phase transition at
$T = T_Q$ as indicated by the transition line.
For $\Gamma=0$, we expect this transition to be in the universality class of the 3D XY model, and the corresponding 
ferroquadrupolar order parameter $m_{\rm FQ} \!=\! \la |\vec M_{\rm FQ}| \ra/N$ indeed
continuously vanishes when we heat above the transition temperature $T_Q \!\approx\! 1.3 J_1$ as seen from Fig.~\ref{gamma_T_fig}(b). When we turn on $\Gamma\neq 0$, the
clock anisotropy suppresses fluctuations and enhances $T_Q$; furthermore, the transition becomes first order, as is confirmed by the increasingly sharp and discontinuous
drop of $m_{\rm FQ}$ across $T_Q$.

\subsection{Impact of nonzero magnetic field}
\begin{figure}[t]
	\includegraphics[width=0.48\textwidth,height=0.26\textwidth]{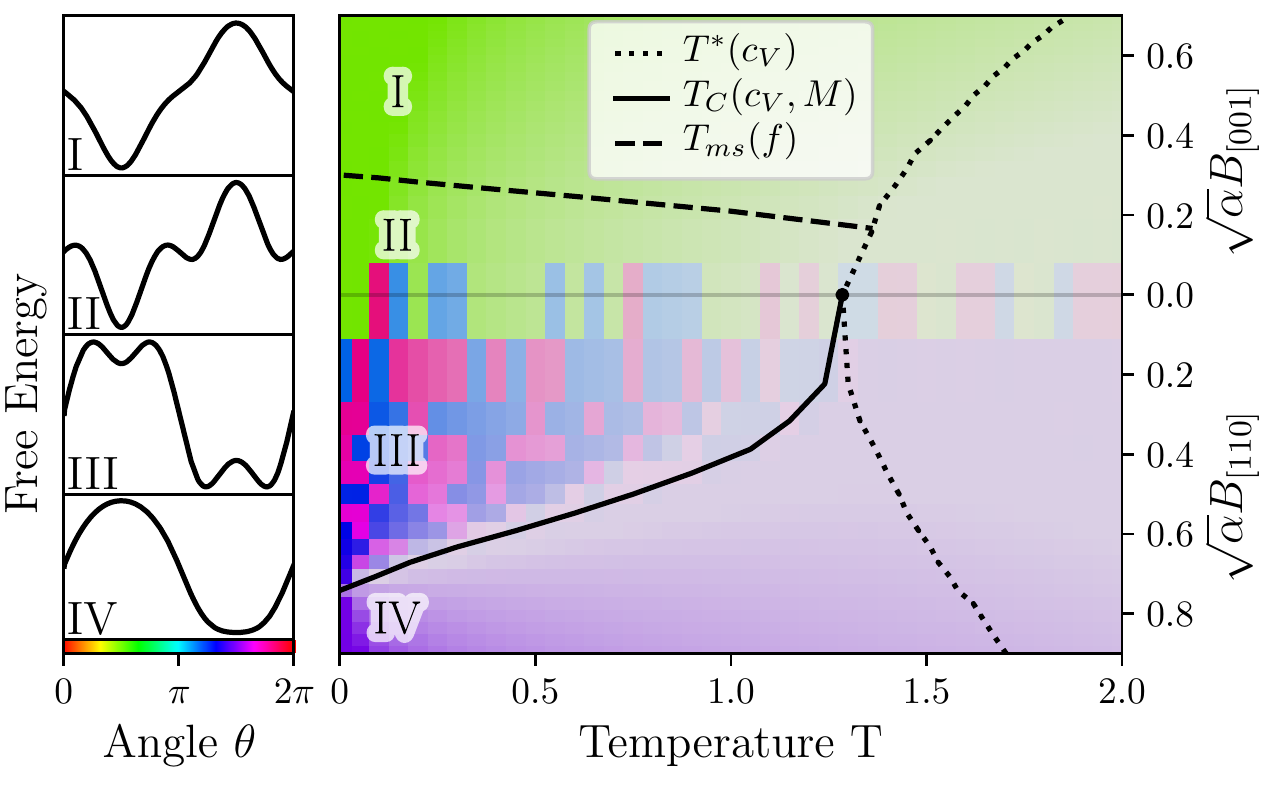}
	\caption{Phase diagram for $B \neq 0$ along the $[001]$ direction (upper half) and $[110]$ direction (lower half) for fixed $J_1\!=\!1$ and $\Gamma\!=\!0.01 J_1$. 
	The left panel indicates the free energy in the different low temperature regimes shown in the phase diagram. The colors in the phase diagram indicate the 
	(dominant) angle $\theta$ as depicted at the bottom of the left panel. Solid line
	shows the Ising phase transition $T_C$ for the $[110]$ field direction which is extracted from specific heat $c_V$ and order parameter $M$, 
	dotted lines depict thermal crossovers $T^*$ obtained from the specific heat $c_V$, and the dashed line shows the
	field where the metastable ($ms$) minima in regime (II) vanish (see left panel) which we extract from free energy ($f$) calculations
	as explained in the text and Appendix.}
	\label{fig:nonzeroB}
\end{figure}

Fig.~\ref{fig:nonzeroB} shows the impact of a magnetic field on the phase diagram for fixed $J_1\!=\!1$ and $\Gamma\!=\!0.01 J_1$. 
The upper and lower halves in this diagram
correspond to fields along the $[001]$ and $[110]$
directions, respectively. We will discuss in the following section that $\Gamma/J_1 \sim 10^{-3}$ for experiments on PrTi$_2$Al$_{20}$; however,
the numerical simulations are more challenging for such small $\Gamma$. We thus choose to work with a larger $\Gamma$ in the MC simulations.
The magnetic field required to induce the relevant transitions or crossovers scales as $\propto \! \sqrt{\Gamma}$ as indicated by mean field theory.
We can thus use our MC results, with suitable scaling, to make useful comparisons with experiment.

As expected, a sufficiently large magnetic field leads to a crossover temperature scale since it favors a single free energy minimum as seen from the
free energy plots for (I) and (IV) in the left panel, where the color at the bottom depicts the favored angle $\theta$.
This crossover temperature $T^*$, indicated by the dotted line, is detected in our MC simulations as a broad hump in the specific heat which does not scale with
system size (based on simulations done for linear system sizes $L=6,9,12$).

At low field, the $[110]$ direction leads to an Ising transition, denoted by the solid black line,
into a phase where there are two degenerate minima as seen from the free energy plot (III) in the left panel. Different MC runs (initialized with a random state) in this regime lead to the system ending up in one or the other minimum, which is depicted by the colors in phase (III) with corresponding $\theta$ values shown in
the left panel.

For the $[001]$ field direction, even at low field, a single free energy minimum is selected as seen from the left panel (II).
The dashed line indicates the crossover field beyond which the metastable
free energy minima in the left panel (II) disappear; this corresponds to the spinodal transition discussed from the perspective of mean-field theory above.

The results from our extensive MC simulations are thus broadly consistent with expectations based on mean field theory, but with thermal fluctuations suppressing the
magnetic field scale required to induce the observed phase transitions and crossovers. We next turn to the experimental implications of this phase diagram.

\section{Experimental implications}

Our classical MC simulations show that the zero field ferroquadrupolar transition for $\Gamma\!=\!0$ occurs at $T_Q \!\approx\! 1.3 J_1$. As
$\Gamma$ increases, $T_Q$ increases and the transition becomes more visibly first order, consistent with the behavior of the 3D $Z_3$ clock
(or equivalently $3$-state Potts) model. Since the experiments \cite{sakai2011kondo,sato2012ferroquadrupolar} 
see what appears to be a nearly continuous thermal transition at $T_Q \! \approx \! 2.2K$,
we assume $\Gamma \! \ll \! J_1$. We thus use the value of $T_Q$ at $\Gamma\!=\! 0$, to roughly estimate $J_1 \!\sim\! 1.7$\,K.
Microscopic calculations \cite{hattori2014antiferro} 
using the measured crystal field levels \cite{sakai2011kondo,sato2012ferroquadrupolar}  
yield $\alpha \!=\! (g \mu_B)^2 (7/3 E_4 - 1/E_5)$, where $g\!=\!4/5$. For PrTi$_2$Al$_{20}$, the
relevant excited crystal field levels \cite{sato2012ferroquadrupolar}  lie at $E_4 \!\approx\! 65$\,K and $E_5 \!\approx\! 107$\,K. This yields $\alpha \approx 0.008$\,K/T$^2$.
Assuming the Ising transition for the $[110]$ field direction \cite{taniguchi2019field, kittaka2019field}
disappears at $B^\star \!\sim\! 3$\,T, we are led to estimate $\Gamma \!=\! \alpha B^{\star 2}/54 \!\approx\! 10^{-3}$\,K, so that indeed $\Gamma \ll J_1$.
The spinodal transition for the $[001]$ field direction is then expected to occur around $B_{\rm sp} \! \sim\! 1.2$\,T. 

Recently, transport, magnetization, and $^{27}$Al nuclear magnetic resonance (NMR) experiments \cite{taniguchi2019field, kittaka2019field} 
have been used to further explore the
phase diagram of PrTi$_2$Al$_{20}$. At high fields, $B \gtrsim 4$\,T, for both [001] and [110] directions, there is a significant enhancement of the
magnetization \cite{kittaka2019field}  upon cooling below $T \!\lesssim\! 3$\,K. However, for low fields, $B \lesssim 2$\,T, this strong enhancement is absent. Within our
theoretical framework, the bulk magnetization is given by the field derivative of the free energy, $\vec m = -\partial F/\partial \vec B$ (note that this is not $m_{\rm FQ}$). 
For both field directions, this is given (up to a sign) by
$|\vec m| \propto B \la \tau_y \ra$. For sufficiently high fields, we expect the system to
evolve from $\la \tau_y \ra \!\sim\! 0$ for high temperature, to a nearly polarized value $|\la \tau_y \ra| \!\sim\! 1$ at low temperature, so that there would be a
significant increase in $|\vec m|$ below a crossover temperature. By contrast, in the presence of metastable domains which we expect at low fields,
$\la \tau_y \ra$ would be greatly reduced via averaging over the domains, since $\la \tau_x \ra$ will also be nonzero in some domains.
This leads to the suppression of the bulk magnetization in low fields, so that the sharp increase upon cooling seen at higher fields will now be absent, in
qualitative agreement with the data.
Furthermore, NMR measurements of the Knight shift \cite{kittaka2019field}  are consistent with
the bulk susceptibility from the magnetization measurements at high field, but in disagreement at low fields; this disagreement might also indirectly signal the presence of
an inhomogeneous domain structure at low fields. The presence of domains is predicted to lead to NMR line splittings, or to inhomogeneous line broadening 
if the splitting is weak. This expectation is qualitatively borne out from the experimental data \cite{kittaka2019field}, but a detailed theoretical understanding
needs further analysis using the microscopic hyperfine couplings. Finally, scattering from such an inhomogeneous domain structure could partially contribute to 
the experimentally observed resistivity anomalies \cite{taniguchi2019field}.

In order to estimate the
typical linear dimension $L_D$ of $Z_3$ domains, 
we ask when the system with an average order parameter pointing along an $XY$ angle $\theta$ would
rather break up into domains of the discrete $Z_3$ order to save bulk anisotropy energy, governed by $\Gamma$,
at the expense of a domain wall cost arising from $J_1$. Assuming a lattice constant $a$,
we thus equate $6 \Gamma (L_D/a)^3 \sim J_1 (L_D/a)^2$ which,
 for $\Gamma/J_1 \sim 10^{-3}$, leads to $L_D \sim 160 a$.
 This might be the size of typical domains we expect to get pinned by elastic lattice deformations.

One possible experimental route to further exploring such a spinodal origin of the magnetization and transport anomalies
could be noise spectroscopy. For instance, resistivity measurements in nanowires of high temperature cuprate superconductors exhibit 
a telegraph noise, which has been attributed to fluctuating nematic domains or charge stripe domains \cite{nematic.1,nematic.2}.
Similar field and temperature dependent resistivity noise measurements might be valuable in PrTi$_2$Al$_{20}$.
Another possible experiment might be to detect the actual sound associated with 
the avalanche of domain rotations one expects near these metamagnetic transitions.

%Using the above numbers, we find that even at $0.8 B_{\rm sp}$ \tcr{$B_{\rm sp}$}, our mean field estimate of the barrier height to be overcome by the metastable minimum is
%$\Delta E \! \approx \! 0.02$ \tcr{0.015}\,K, so that domains of linear dimension \tcr{$L \!\gtrsim\! (T_Q/\Delta E)^{1/3}\! \approx\! 5$} lattice constants are expected to be pinned.
%The pinning and energy barriers would however be strongly dependent on the coupling to the phonons, which would quantitatively impact
%the location of the proposed spinodal transitions.

Our proposal of strong spin-lattice coupling leading to field-induced anomalies is distinct from, but not entirely at
odds with, a previously proposed explanation \cite{kittaka2019field}, which has considered the impact of additional field-dependent quadrupolar exchange couplings
within an effective Landau theory. While the microscopic origin of this effect has been attributed to field-induced changes in the Fermi surface \cite{kittaka2019field}, and
thereby the RKKY Kondo couplings, such terms may also occur if we incorporate field dependent spin-phonon coupling and integrate out the phonons.
The microscopic details of such a mechanism, and its connection with the metastable domain picture discussed here, remains a 
topic for future study.

In summary, understanding the nature of the field-dependent phase transitions and anomalies in PrTi$_2$Al$_{20}$
may help deepen our understanding of multipolar orders in heavy fermion materials. Finally,
our work suggests that multispin interactions must play a broadly important role in Kondo materials.

\acknowledgments

A.P. acknowledges funding from NSERC of Canada. 
S.T. and A.P. acknowledge
partial funding from the DFG within CRC
1238 (project C02), Projektnummer 277146847.
Y.B.K. was supported by the Killam Research Fellowship from the Canada Council 
for the Arts and NSERC of Canada.  S.B.L. is supported
by the KAIST startup and National Research
Foundation Grant (NRF-2017R1A2B4008097). 
%The numerical simulations were performed on the JUWELS
%cluster at FZ Juelich. 
The numerical simulations were performed on
the CHEOPS cluster at RRZK Cologne.
F.F. thanks the Bonn-Cologne Graduate School of Physics and Astronomy
(BCGS) for support.

\bibliography{multipolar}

\appendix

\section{Algorithms}

Equilibrium states of classical many-body systems can be probed by standard Monte-Carlo simulations. For system, which do not allow for efficient non-local (cluster) update, extended ensemble approaches such as simulated annealing and parallel tempering are often used to find the equilibrium state more efficiently and accurately. If one is, however, primarily interested in {\em metastable} states, then non-standard procedures are often called for. In the following sections, we describe our problem-specific approach of resolving metastable states for model \eqref{eq:H0}. Notably, our approach also allows us to probe the free energy as function of angle $\theta$.

%\subsection{Half-sphere sampling}
%
%\textcolor{red}{Are we keeping plots from these simulations?}
%
%One idea we applied is to simply exclude the dominant state from our simulations. Given that this state is a ferromagnetic one, we simply drop a certain range of spins from our local update. Instead of sampling new spins $S = (\cos(\theta), \sin(\theta))^T$ with $\theta \in [0, 2\pi]$, we sample $\theta \in [\pi, 2\pi]$.
%
%This update is strongly biased. For example, it completely disallows Neél order in with $\theta = \pm \pi/2$.

% tl;dr
% - sample only spins (cos(theta), sin(theta)) where theta in [180°, 360°]
% Bias: just the obvious - theta in [0, 180] excluded

\subsection{Pair-sampling}

We start by discussing a tailor-made update procedure for our model, which allows us to simultaneously sample two dominant angles of the XY order parameter
and, by recording the number of sweeps spent on each, the relative probability between them. The procedure is split into two parts, a local update which preserves the 
XY order parameter angle $\theta$ and global update which perform jumps between the two angles of interest. The latter is simply a global update rotating the whole spin configuration back and forth. The former is more complex and requires a more thorough discussion. As the local update is explicitly biased (by forcing $\theta$ to be constant) we present two algorithms and briefly discuss the effect of different biases on the derived free energy.

\begin{figure}[h]
	\includegraphics[width=\columnwidth]{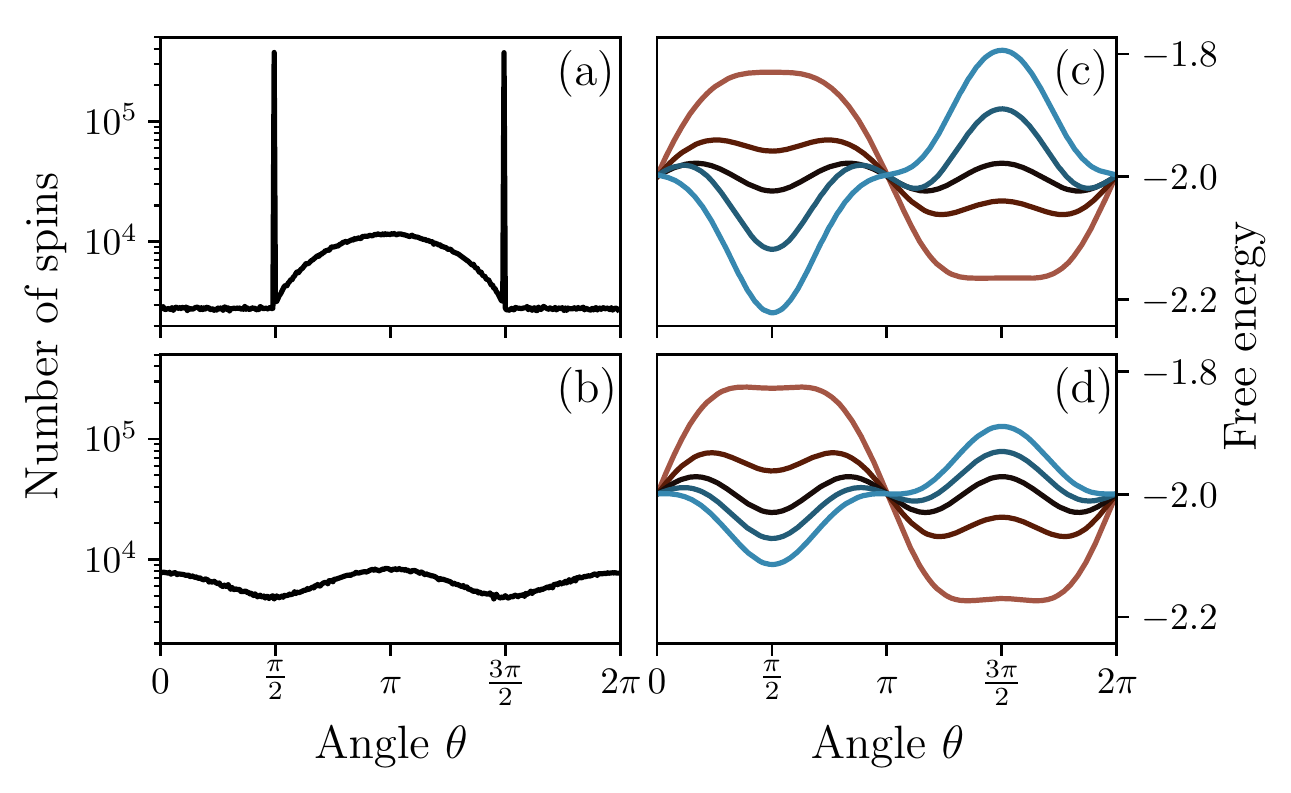}
	\caption{Comparison between pair-sampling algorithms. The different algorithms are shown in rows, with the top row (a, c) being the first algorithm and the bottom row (b, d) being the second; the temperature $T=J_1$ for this plot.
	The left column (a, b) shows a histogram of the proposed spins, where $\hat{e}_M = (\cos(\pi), \sin(\pi))$ is set for both. The right column (c, d) shows the free energy curves resulting from the respective updates for magnetic fields $\sqrt{\alpha} B \approx 0, 0.28, 0.49$ in [110] and $\sqrt{\alpha} B \approx 0.2, 0.32$ in [001] direction at $T=1$.}
    \label{fig:sampling}
\end{figure}

Just for the purpose of this Appendix, we introduce slightly convenient notation, denoting the unit vector $\vec S_i \equiv \vec\tau^\perp_i$. 
The XY order parameter in a given configuration is then
as $\vec M_{\rm FQ} = \sum_i \vec S_i$, and
$\hat{e}_M = \vec M_{\rm FQ}/|\vec M_{\rm FQ}|$ will be the direction of the XY order parameter, 
and $\hat{e}_\perp = \hat{e}_z \times \hat{e}_M$ denotes 
the direction perpendicular to it. If we want to sample configurations with fixed $\hat{e}_M$, i.e. a fixed angle in which the global XY order
parameter points, we must use
configuration updates $\{ \vec S^\pdg_i \} \to \{S^\prime_i \}$ such that
$\sum_i (\vec S^\prime_i - \vec S_i) \cdot \hat{e}_\perp = 0$.
The first algorithm proceeds to do this in following steps:
\begin{enumerate}
	\itemsep0em
	\item Pick a random site $i$ and a new random spin $\vec S_i^\prime$. Compute $x =  (\vec S^\prime_i - \vec S_i)\cdot \hat{e}_\perp$ 
	which is the component in the $\hat{e}_\perp$ direction that 
	must be compensated. %For simplicity let us assume that $x > 0$, so that we need a compensation in $-e_\perp$ direction.
	\item Pick a random site $j$ which has not yet been chosen. Compute $\Delta x = (\vec S^\prime_j - \vec S_j)\cdot \hat{e}_\perp = \mp 1 - \vec S_j \cdot \hat{e}_\perp$ which is the maximum compensation that
	can be achieved by setting $\vec S_j \to \mp \hat{e}_\perp$ (respectively for $x \gtrless 0$).
	\item If $x^\prime = x - \Delta x = 0$  or changes sign, the random spin flip can be fully compensated. Compute the necessary $\vec S_j^\prime$ 
	and return every changed spin as a proposed update.
	\item If $x^\prime$ has the same sign as $x$ the random spin flip cannot be compensated. Set $\vec S_j \to \mp \hat{e}_\perp$, $x \to x^\prime$ and go to (2), picking an additional (unique) site for the update.
\end{enumerate}
We note that this algorithm comes with a strong, localized bias because updates frequently include setting one or more spins $\vec S \to \pm \hat{e}_\perp$. Fig.~\ref{fig:sampling}(a) shows a histogram of the proposed spins, making the bias obvious as two sharp peaks. The second algorithm is designed to avoid this bias. It includes the following steps:
\begin{enumerate}
	\itemsep0em
	\item Pick spins at two distinct random sites $\{i_1, i_2\}$ and two new random spin vectors $\{\vec S_1^\prime, \vec S_2^\prime\}$.
	\item Compute combined vector length $a = |\vec S^\prime_1 + \vec S^\prime_2|$ and combined component $b = (\vec S_{i_1} + \vec S_{i_2}) \cdot \hat{e}_\perp$ that must be compensated.
	\item If $a > b$ a rotation $R$ can be found, such that $\{R \vec S_1^\prime, R \vec S_2^\prime, \dots\}$ keeps the XY order parameter direction $\hat{e}_M$ 
	unchanged. Compute this rotation and return the rotated spins as a proposed update.
	\item If $a < b$ we cannot find such a rotation. Add a new random spin $\vec S^\prime_3$ and a 
	new (unique) random site $i_3$ to the collection of updated spins and sites and repeat from step (2).
\end{enumerate}
This update is weakly biased in $\pm \hat{e}_M$ direction. As evident in Fig.~\ref{fig:sampling}(b) the distribution of proposed spins is much smoother. 

% tl;dr - Algorithm 1 (used for everything)
% 1. pick random spin, random site (S' at i)
% 2. compute change perpendicular to M: x = \langle S' - S, e_z \times e_M\rangle
% 3. pick random site j (not picked before)
% 4. compute maximum compensation possible by rotating site j to \pm M_\perp
% 5. if it cannot compensate S -> S':
%	- rotate spin at site j to \pm M_\perp
% 	- update x -> x - max_compensation
%   - goto 3
% 6. Rotate spin at site j to exactly compensate x
% Bias: strong in \pm e_z \times e_M direction (perpendicular to M)

% tl;dr Algorithm 2 (checked 1 against this)
% 1. pick 2 random spins, 2 random sites
% 2. if ||\sum_i S_i'|| > \langle \sum_i S_i, e_z \times e_M\rangle
%	- rotate all S_i' so that
%		\langle \sum_i S_i' - S_i, e_z \times e_M\rangle = 0
% 3. if number of new spins > 5
%	- goto 1
% 4. pick one more random spin and site, goto 2
% Bias: weak, gradual bias towards \pm M direction

\subsection{Computation of Free Energy}

The pair-sampling method allows us to fix two angles of the XY order parameter $\{\theta, \theta + \Delta\theta\}$. By counting the number of sweeps spent at each angle we can determine the relative weight between them
\begin{equation}
	\frac{Z(\theta + \Delta\theta)}{Z(\theta)} = \frac{N(\theta + \Delta\theta)}{N(\theta)} \,.
\end{equation}
By setting an initial value for $Z(0)$ we can compute successive $Z(\theta > 0)$. From this we can derive the angle resolved free energy
$F(\theta) = -\log(Z(\theta)) / \beta$.
Note that this process becomes increasing expensive at low temperatures, requiring small $\Delta\theta$ and a large number of sweeps to get finite counts $N(\theta) > 0$.
The free energy per site $F(\theta)/N_{\rm site}$ from the two algorithms is compared in Fig.~\ref{fig:sampling}(c) and \ref{fig:sampling}(d), and they show 
very similar angle dependence, although there is some difference is in the amplitude of the free energy variation for a field applied along the
[001] direction. The second algorithm with a smoother distribution of proposed updates is likely to be a better representation of the true free energy curve. 
Comparing the computed result to the mean-field free energy, we find that the angle
dependence is nearly identical; the Monte Carlo and mean-field curves match closely up to an overall $\sim 2.4$ scale factor, which reflects a renormalization
of $\Gamma$ due to thermal fluctuation effects beyond mean field theory.

%In practice we need to make sure that this can actually be computed. To get a finite count $N(\theta) > 0$ for all $\theta$, $\Delta\theta$ must be sufficiently small and the number of sweeps sufficiently large. This becomes increasingly difficult as $T \to 0$, limiting our ability probe low temperature behavior.

\end{document}